\documentclass[aps,prl,reprint,showpacs,showkeys,superscriptaddress]{revtex4-1}
\usepackage[pdftex]{hyperref,color,graphicx}
\hypersetup{colorlinks=true,linkcolor=blue,citecolor=blue,filecolor=blue,urlcolor=blue}
\usepackage{amsfonts,amssymb,amsmath}
\usepackage[english]{babel}

\newcommand*{\ket}[1]{|{#1}\rangle}
\newcommand{\figref}[2]{\hyperref[#1]{\ref{#1}(#2)}}
\newcommand{\definition}{:\kern-0.3pt=}
\makeatletter
\renewcommand*{\fnum@figure} {FIG. \thefigure{} (color online)}
\makeatother

\usepackage{textcomp}

\begin{document}
\selectlanguage{english}

\title{
Electric quantum walks with individual atoms
}

\author{Maximilian Genske}
\author{Wolfgang Alt}
\author{Andreas Steffen}
\affiliation{Institut f\"ur Angewandte Physik, Universit\"at Bonn, Wegelerstra\ss{}e 8, D-53115 Bonn, Germany}
\homepage{http://quantum-technologies.iap.uni-bonn.de}
\email{alberti@iap.uni-bonn.de}
\author{Albert H. Werner}
\author{Reinhard F. Werner}
\affiliation{Institut f\"ur Theoretische Physik, Leibniz Universit\"at Hannover, Appelstrasse 2, 30167 Hannover, Germany}
\author{Dieter Meschede}
\author{Andrea Alberti}
\affiliation{Institut f\"ur Angewandte Physik, Universit\"at Bonn, Wegelerstra\ss{}e 8, D-53115 Bonn, Germany}

% \date{\today}

\pacs{
	05.60.Gg, % Quantum transport
	03.75.-b, % Matter waves
	73.40.Gk  % Resonant tunneling
}

\keywords{Quantum walks; Bloch oscillations; Floquet theory; Quantum resonances}
 
\begin{abstract}
	We report on the experimental realization of electric quantum walks, which mimic the effect of an electric field on a charged particle in a lattice. Starting from a textbook implementation of discrete-time quantum walks, we introduce an extra operation in each step to implement the effect of the field. The recorded dynamics of such a quantum particle exhibits features closely related to Bloch oscillations and interband tunneling. In particular, we explore the regime of strong fields, demonstrating contrasting quantum behaviors: quantum resonances vs.\ dynamical localization depending on whether the accumulated Bloch phase is a rational or irrational fraction of $2\pi$.
\end{abstract}

\maketitle

%
% Word counting.
%
% Abstract: 98
% Body+captions: 2560
% Figure1: 160 equivalent words
% Figure2: 257 equivalent words
% Figure3: 129 equivalent words
% Figure4: 97 equivalent words
% Total figures: 643 equivalent words
% Grandtotal: 3200/3500 words

Discrete-time quantum walks conceptually represent the simplest realization of transport in a quantum system. In essence, a spin-1/2 particle moves on a lattice in discrete steps, with the direction determined by its internal spin state. A coin operation acting on the internal states is applied at each step, allowing for the coherent coupling of spin-dependent quantum paths.
The iteration of the shift and coin operation delocalizes the particle over a complex host of paths that interfere with each other, determining the relevant transport properties. Emblematic of quantum walks is, for instance, the ballistic spreading of the walker, which contrasts strikingly with the diffusive classical transport of random walks. For a review, see \cite{VenegasAndraca:2012} and references therein.

This quantum transport model, first introduced by Feynman while developing a path integral formulation of quantum mechanics \cite{FeynmanBook}, still holds great relevance because it is simple and yet very powerful. In fact, quantum walks have recently been shown to constitute a universal computational primitive \cite{Childs:2009} and to provide the basis for a series of quantum algorithms \cite{Shenvi:2003}. Over the past few years, it has become technologically possible to implement discrete-time quantum walks in real systems and to realize the coin operation in a variety of ways: using cold atoms \cite{Karski:2009} and trapped ions \cite{Schmitz:2009,Zahringer:2010} with the qubit encoded in hyperfine states, or using single photons  with the qubit encoded in either polarization states \cite{Broome:2010} or different spatial modes \cite{Sansoni:2012}.

Since their first experimental realization, discrete-time quantum walks, as well as the closely related continuous-time versions, have yielded numerous achievements: to mention only a few, quantum correlations between identical walkers \cite{Peruzzo:2010}, observation of topological protected states \cite{Kitagawa:2012}, and the prediction of artificial molecular states \cite{Ahlbrecht:2012}. However, little attention has been paid by experiments to the connection between the behavior of quantum walks subject to an external driving field and the underlying dispersion relations. In this respect, we present in this Letter an experimental study of the dynamics of a single quantum particle performing a one-dimensional quantum walk under the application of an artificial electric field. The principal aspects of this problem have been theoretically addressed, like the recurrence probability as a function of the field strength \cite{Wojcik:2004,*Banuls2006}. Very recently we have mathematically studied in detail the spectral properties of the system, which provide guidance to experiments in the long-time limit \cite{Werner2012}. Some of these aspects have been also reproduced with a classical analogue of quantum walks, employing laser pulses circulating in a fiber loop network~\cite{Regensburger:2011}.

\begin{figure}[!t]
	\centering
		\includegraphics{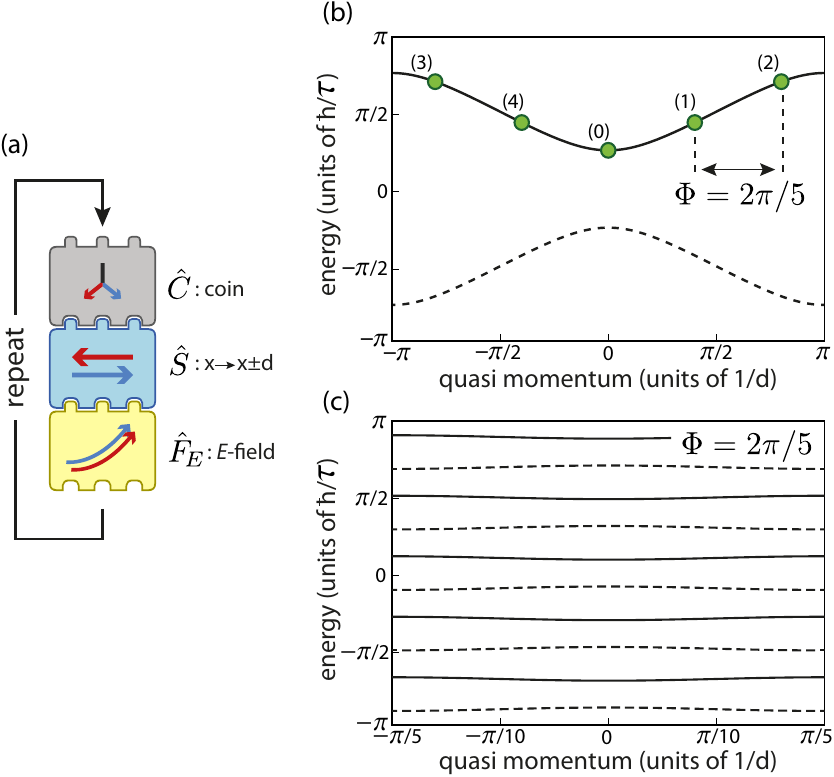}
	\caption{(a) Block representation of an electric quantum walk. (b) The quasi momentum shifts in discrete units of $\Phi/d$ at each step within the Brillouin zone. Solid and dashed curves are the two energy bands of a quantum walk in zero field. Starting for example with $k=0$, the momentum returns to its initial value after 5 steps, similar to Bloch oscillations (interband tunneling is here omitted). (c)  The energy states of a rational electric quantum walk form nearly flat bands, resembling two Wannier-Stark ladders (solid and dashed lines).}
	\label{fig:Figure1}
\end{figure}

In a discrete-time quantum walk, quantum states evolve by applying for each step the unitary operator $\hat{W}_0\definition\hat{S}\hspace{0.3pt}\hat{C}$, where $\hat{C}$ is the coin and $\hat{S}$ is the shift operator. The walk can be turned into an electric one by adding the extra operation $\hat{F}_E\definition\exp(i\hspace{0.3pt}\Phi\hspace{0.3pt}\hat{x})$ to its sequence, with $\hat{x}$ being the lattice position operator. Here $\Phi$ represents the phase imprinted by the effective electric field between two adjacent lattice sites at each step.  In other words, this phase represents the action of an electric field $E$ coupled to a charge $q$, which is justified by the formal definition $\Phi\definition{}q\hspace{0.3pt}E\hspace{0.3pt}\hspace{0.3pt}d\hspace{0.3pt}\tau/\hbar $, with $d$ being the lattice constant and $\tau$ the duration of a single step of the walk. A block illustration of the resulting electric walk, $\hat{W}_E\definition\hat{F}_E\hspace{0.3pt}\hat{W}_0$, is displayed in Fig.~\figref{fig:Figure1}{a}. Because the zero-field walk $\hat{W}_0$ is translationally invariant, the natural way to study the dynamics of the system in the presence of an electric field is to use the Fourier representation where the quasi momentum $\hat{k}$ is a diagonal operator. In $k$-space, the operations constituting $\hat{W}_0$ can be conveniently expressed in terms of Pauli matrices: $\hat{S}:\kern-0.3pt=\exp(-i\hspace{0.3pt}k\hspace{0.3pt}d\hspace{0.3pt}\hat{\sigma}_z)$ is the spin-dependent shift and $\hat{C}:\kern-0.3pt=\exp(-i\hspace{0.3pt}\pi\hspace{0.3pt}\hat{\sigma}_y/4)$ is the Hadamard coin. Owing to two internal states, the energy spectrum of $\hat{W}_0$ posses two bands, see Fig.~\figref{fig:Figure1}{b}, which are functions of $k$ in the Brillouin zone $[-\pi/d,\pi/d]$. In this representation, the field operator can be written as $\hat{F}_E=\exp[-(\Phi/d)\hspace{0.6pt}\partial/\partial k]$, which represents a displacement of $k$ by $\Phi/d$ modulo the Brillouin zone, as indicated in Fig.~\figref{fig:Figure1}{b}. If $\Phi$, which we call Bloch phase, is commensurable with $2\pi$, i.e., $\Phi=2\pi\,n/m$ with $n$ and $m$ without a common factor, after $m$ steps the quasi momentum returns to its original position. We will call this a ``rational electric field'' to distinguish it from an ``irrational'' one.

The dynamics of this system exhibit a close analogy with the problem of Bloch oscillations of a charged particle in an electric field, which has been used in other systems as a hallmark of quantum transport: using cold atoms \cite{BenDahan:1996}, photons \cite{Pertsch:1999,*Morandotti:1999,*Sapienza:2003}, and electrons in superlattices \cite{Feldmann:1992,*Leo1992}. Beyond this evident similarity, the discrete-time nature of the system gives rise to unique transport features, which reflect the commensurability of the Bloch phase with $2\pi$.
For instance, with a rational electric field, one realizes that the system is still invariant under a translation of a multiple of $m$ sites. This is in striking contrast to the continuous-time situation of an electron in a lattice under the action of an external electric field, where the lack of translational symmetry leads to Wannier-Stark localization \cite{Gluck:2002}.
In the discrete-time case instead, this translational symmetry allows us to define a quasi momentum $k$ in a restricted Brillouin zone $[-\pi/(md),\pi/(md)]$, producing the energy bands depicted in Fig.~\figref{fig:Figure1}{c}.
Their analytic expression is given in \cite{Werner2012}. 
In general, the singular properties of the discrete-time case become more pronounced in the regime of strong fields, which is defined by Bloch phases that are a sizable fraction of $2\pi$.
We will focus in this Letter explicitly on this regime, in order to put in clear evidence the new properties of electric quantum walks.

\begin{figure}[!t]
	\centering
		\includegraphics{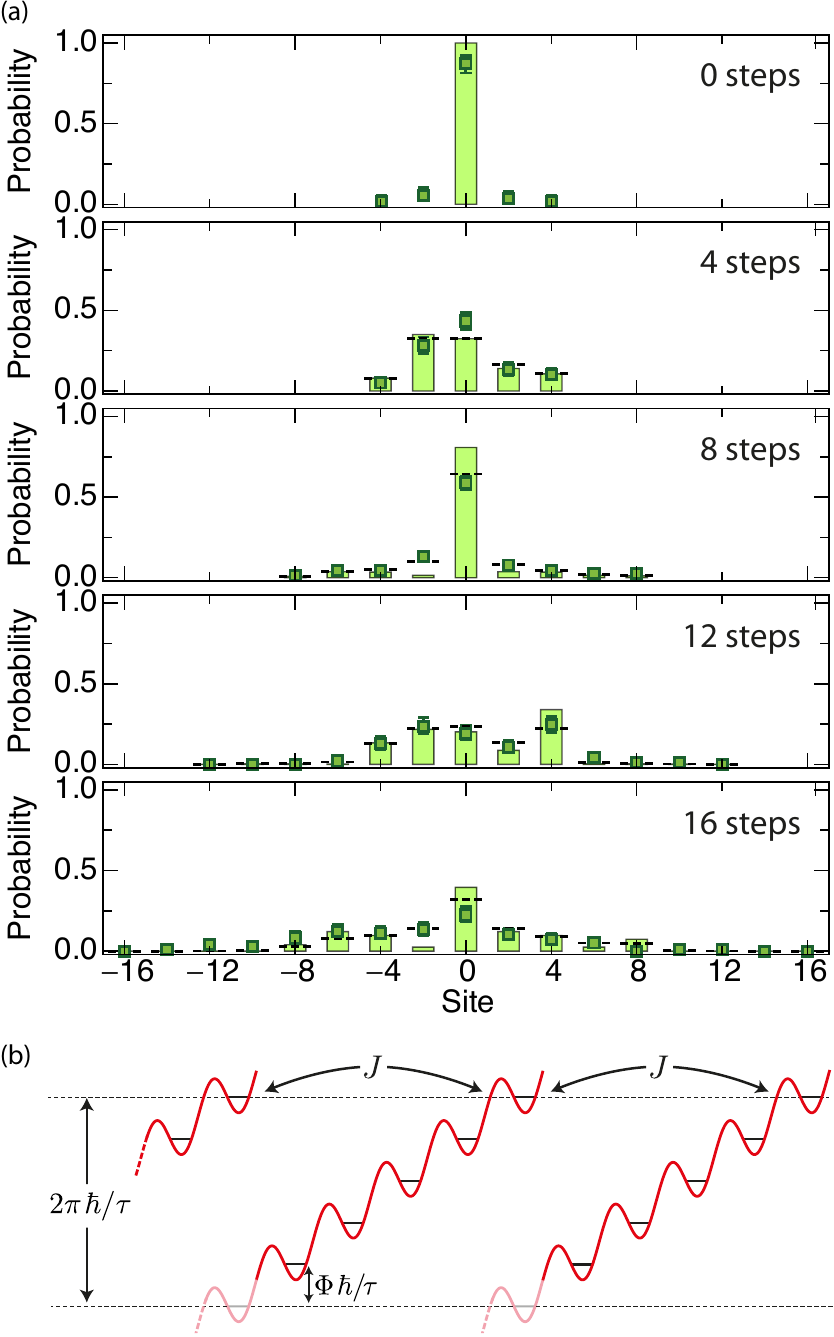}
	\caption{(a) Bloch-oscillation-like revivals in an electric quantum walk. Partial revivals at $8$ and $16$ steps show a clear evidence of resonant tunneling for the fractional electric field $\Phi=2\pi/8$. The vertical bars indicate the theoretical distribution, the dashed horizontal lines the theoretical distribution corrected for spin dephasing ($10\%$ per step), and the square points with error bars (68\% Clopper-Pearson confidence intervals) the measured values. (b) Floquet theory provides an intuitive picture of the resonant tunneling mechanism for rational electric fields (in the sketch $\Phi=2\pi/5$).}
	\label{fig:Figure2}
\end{figure}

Discrete-time quantum walks are implemented in our systems using single Cs atoms in spin-dependent optical lattices: the pseudo-spin-1/2 state $\ket{\uparrow}=\ket{F=4,m_F=4}$ or  $\ket{\downarrow}=\ket{F=3,m_F=3}$ is trapped in its respective periodic potential, created by  $\sigma^{+}$\hspace{-1pt}- or $\sigma^{-}$\hspace{-1pt}-polarized optical standing waves.
In each lattice well, atoms are cooled to the motional ground state in the lattice direction by means of microwave sideband cooling, allowing a spin coherence time of $300$\,\textmu{}s.
Because the lattice potential is very deep (800 recoil energy units), each spin component of the atom strictly follows the displacement of the corresponding spin-dependent lattice, without undergoing any tunneling between sites.
In this sense, during a spin-dependent shift operation each spin component moves along well-defined trajectories. Each shift lasts $24$\,\textmu{}s, which is chosen to minimize motional excitations inside the potential wells to below $1\%$. The relative displacement between the two lattices is varied with nanometer precision from $0$ to $d$ at one step and from $d$ to $0$ at the next step, where in our system $d=433\,$nm \cite{Karski:2009}; it can be shown that the resulting walk is fully equivalent to the one represented in Fig.~\figref{fig:Figure1}{a}.
The coin operation is implemented by microwave pulses lasting about $11$\,\textmu{}s, which realize a $3\pi/2$ rotation rather than a $\pi/2$ rotation to take advantage of a partial refocussing of inhomogeneous dephasing effects.
The phase $\Phi$ is controlled by accelerating both spin-dependent lattices in the same direction for a certain time at each step, producing in the reference frame of the lattice an identical inertial force on both spin states.
The acceleration is realized by quadratically ramping the phase of one lattice arm with a direct digital synthesizer; $\Phi$ is independently measured by using a two-site-splitting atom interferometer \cite{Steffen:2012}.
Both the spin-dependent shift and the lattice acceleration leave the atom in the motional ground state with a probability higher than $99\%$. At each run of the walk, the displacement of the atom is determined by measuring its position in the lattice before and after the walk by fluorescence imaging, with an efficiency at around $90\%$ to retrieve the exact lattice site.

\newsavebox{\circlesymb}
\newsavebox{\trianglesymb}
\savebox{\circlesymb}{\begin{picture}(5,5)\put(2.5,2.5){\circle*{4}}\end{picture}}
\savebox{\trianglesymb}{\begin{picture}(5,5)\put(2.5,2.5){\circle*{4}}\end{picture}}
\begin{figure}[!b]
	\centering
		\includegraphics{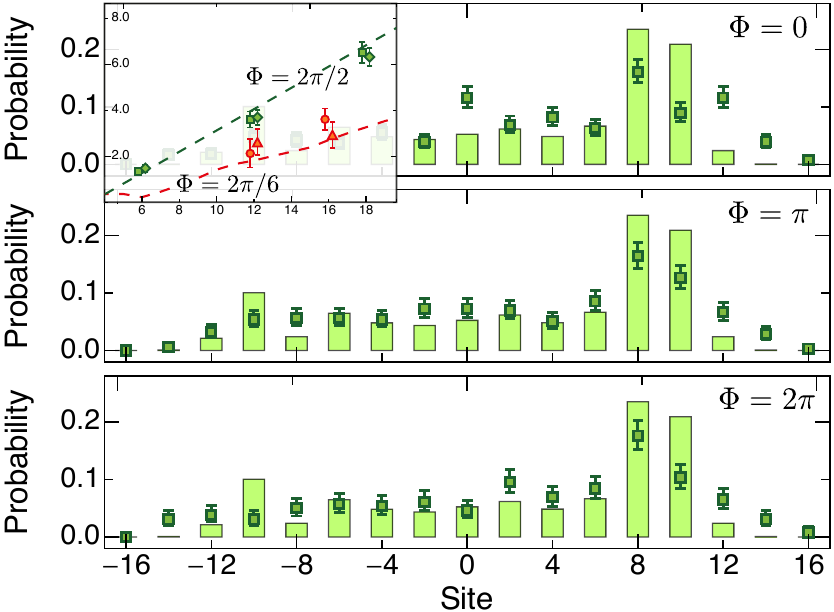}
	\caption{The identical expansion in the three cases provides an indirect signature of an interband tunneling entirely swapping the bands' populations. The distribution refer to 18-step walks.  Inset: \protect\raisebox{0pt}[0pt][0pt]{$\langle{}x^2\rangle{}^{1/2}$} widths comparing the case $m=3$	(\protect\scalebox{1.1}{$\bullet$}) with $m=6$ (\protect\scalebox{0.95}{$\blacktriangle$}) and $m=1$ (\protect\rule{4pt}{4pt}) with $m=2$ (\protect\rotatebox{45}{\protect\rule{4pt}{4pt}}) ($n=1$); the dashed lines are the theoretical predictions.}
	\label{fig:Figure3}
\end{figure}

In order to study the dynamics of an electric quantum walk in the regime of strong rational fields, we prepare a single atom in a given lattice site with spin $\ket{\uparrow}$ and we let it evolve for an increasing number of steps of the sequence described in Fig.~\figref{fig:Figure1}{a} with $\Phi=2\pi/8$. In the physical picture presented in Fig.~\figref{fig:Figure1}{b}, the initial state corresponds to an equal superposition of all momentum states, with the relative weight between the two bands as a function of $k$. For this value of $\Phi$, one expects a revival of the initial distribution for every $8$ steps, i.e., a peak at the origin, due to an integer number of cycles of $k$ around the Brillouin zone. The measured distributions in Fig.~\figref{fig:Figure2}{a} indeed exhibit evidence of revivals at $8$ and $16$ steps. However, neither the measurements nor the theoretical values show a full revival signal, but rather a spreading behavior, which becomes ballistic for large number of steps.
The origin of this behavior lies in the small yet non-negligible band curvature in Fig.~\figref{fig:Figure1}{c}. The Floquet theory helps us construct an intuitive picture of this mechanism, where we represent the energy states on a periodic potential tilted due to the external field and folded in the interval $2\pi\hbar/\tau$, see Fig.~\figref{fig:Figure2}{b}; in fact, the eigenvalues of $\hat{W}_E$ allow energy to be defined only up to a multiple of this interval \cite{Shirley:1965}. It becomes clear then that at every $m$ sites the energy levels are resonantly aligned allowing for certain tunneling $J$, which corresponds to the curvature of the bands in Fig~\figref{fig:Figure1}{c}.
The same mechanism is interpreted in terms of quantum resonance in quantum kicked rotor systems \cite{Sadgrove:2005,*Ryu:2006} and in modulated optical lattices \cite{Ivanov:2008,*Sias:2008}.

It is of interest to connect the observed behavior of quantum resonances to interband tunneling. While every step shifts the quasi momentum $k$, the internal state remains unaffected by $\hat{F}_{E}$, thus resulting in a tunneling between the two bands in Fig.~\figref{fig:Figure1}{b}. We choose to focus on the strong driving regime where the typical adiabatic assumption to treat the tunneling no longer applies. The analytical solution of the problem predicts for $\Phi=\pi$ a dramatic diabatic tunneling that entirely swaps the populations of the two bands at every step. The distribution recorded for this case is compared to the case of $\Phi=2\pi$, which in turn is expected to be equivalent to the case of no force, see Fig.~\ref{fig:Figure3}. This comparison reveals a striking resemblance of the three cases. This can be understood by realizing that, for any momentum state $k$, the simultaneous momentum shift by $\pi/d$ and band swap leaves the first derivative of the energy band unchanged, i.e., the group velocity. Because in general the asymptotic distributions are only determined by the group velocities, it implies that the three fields must produce the same asymptotic dynamics. Such an experimental agreement, hence, provides an indirect evidence of a ``super interband tunneling'' which fully reverses the bands' populations at every step. In addition, similar arguments based on interband tunneling reveal that this singular  property of identical asymptotic dynamics applies to all pairs $\Phi=2\pi\,n/m$ and $\Phi=2\pi\,n/(2m)$ with $m$ being an odd number. This is experimentally confirmed by comparing the distribution widths of the pairs $m=(3,6)$ and $m=(1,2)$ with $n$ always set to 1, as shown in the inset of Fig.~\ref{fig:Figure3}.

\begin{figure}[!t]
	\centering
		\includegraphics{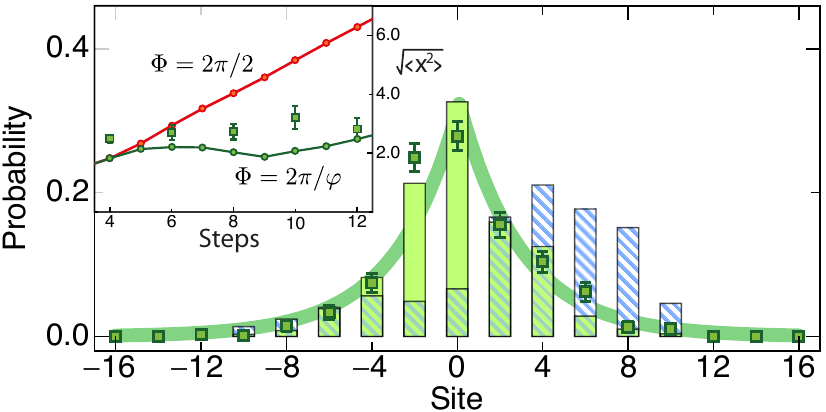}
	\caption{Localized distribution in the case of an irrational electric field, $\Phi=2\pi/\varphi$ with $\varphi$ being the golden ratio. The square points represent the average over a series of  distributions measured at 4, 6, 8, 10 and 12 steps, which are compared to expected values (vertical green bars); the thick line is an exponential curve to guide the eye. In contrast, the overlaid hatched bars show the expected values for $\Phi$ equal to $\pi$ instead of $2\pi/\varphi$. Inset: \protect\raisebox{0pt}[0pt][0pt]{$\langle{}x^2\rangle{}^{1/2}$} widths of the measured  distributions (square points) and of the theoretical distributions for the two values of the electric field (round points).}
	\label{fig:Figure4}
\end{figure}

In contrast, the transport dynamics produced by an irrational electric field are substantially different. In this case, dynamical localization is conjectured for most values of the irrational field, meaning that even after arbitrarily many time steps the quantum walker remains confined to a finite region of the lattice, up to exponentially small corrections \cite{Werner2012}. In order to provide experimental evidence for this fact, we choose $\Phi=2\pi/\varphi$ as the electric field with $\varphi=(\sqrt5+1)/2=1.618\ldots$ being the golden ratio.
We investigate the dynamics of an atom that is initially prepared in a single lattice site with spin $\ket{\uparrow}$. In the case of localization, the walk is expected to remain within a limited region and it is therefore natural to study the time-averaged distribution. Recording several distributions with an increasing number of steps produces the result shown in Fig.~\ref{fig:Figure4}, which is in clear agreement with the expected theoretical prediction. Furthermore, the shape of the distribution closely matches a two-sided exponential profile, which can be interpreted as a signature of the conjectured dynamical localization.
The observed difference between rational and irrational fields raises the natural question: do irrational numbers exist in nature? The immediate answer is that no experiment can distinguish between rational or irrational numbers. However, irrational numbers can be approximated with increasing precision. In oder to discriminate between two fields which differ by $\delta\Phi$, one needs a number of steps on the order of $N\sim1/{\delta\Phi}$.
For example, Fig.~\ref{fig:Figure4} shows a visible discrepancy between the measured distribution and the expected one for the case $\Phi=\pi$.

In conclusion, we have presented an extensive experimental study of the transport dynamics of  electric quantum walks. This work confirms several theoretical predictions, ranging from quantum resonances for rational fields to the conjectured dynamical localization for irrational ones \cite{Wojcik:2004,Werner2012}. Extending the coherence time of the system, by cooling the atoms transversally to the ground state for instance, would permit longer sequences, which are necessary to investigate the localization properties of the wave packets for irrational fields. In particular, one would expect to observe a self-similarity occurring in the energy spectrum, which resembles the Hofstadter butterfly \cite{Artoni:2009}. Preparing initial states with given momentum $k$ would provide direct experimental access to dispersion relations and an alternative method to explore the energy spectrum; this seems experimentally feasible by either better cooling or simply employing a filtering scheme. Furthermore, this experiment opens the way towards the simulation of Maxwell's equations in discrete-time systems. We deem it possible to implement artificial magnetic fields by extending the present apparatus to two-dimensional, spin-dependent optical lattices \footnote{In preparation}. More importantly, this should allow us to access the strong field regime characterized by Peierls phases comparable with $2\pi$.

\begin{acknowledgments}  The authors are indebted to E.~Arimondo for critical reading of the manuscript. We acknowledge financial support from DFG Research Unit FOR 635, NRW-Nachwuchsforschergruppe ``Quantenkontrolle auf der Nanoskala'', ERC grant DQSIM, EU project AQUTE, and Studienstiftung des deutschen Volkes. AA and MG also acknowledge support from the Alexander von Humboldt Foundation and the BCGS, respectively.  

\end{acknowledgments}

\bibliography{bibliography}

\end{document}